\documentclass[aps,prb,superscriptaddress,longbibliography,twocolumn]{revtex4-2}

\usepackage[T1]{fontenc}
\usepackage{lmodern}
\usepackage[svgnames]{xcolor}
\usepackage{lipsum}
\usepackage{graphicx}
\usepackage{bbold}

\definecolor{Dark}{gray}{0.2}
\definecolor{MedDark}{gray}{0.4}
\definecolor{Medium}{gray}{0.6}
\definecolor{Light}{gray}{0.8}
\definecolor{darkred}{rgb}{0.55, 0.0, 0.0}
\definecolor{darkslateblue}{rgb}{0.28, 0.24, 0.55}
\definecolor{royalblue(web)}{rgb}{0.25, 0.41, 0.88}
\usepackage{graphicx,float}
\usepackage{amsmath}

\usepackage{amssymb}
\usepackage[colorlinks]{hyperref}
\hypersetup{
    citecolor=darkred,
    linkcolor=blue,   
    urlcolor=blue}
    
\usepackage{bbm}

\def\be{\begin{equation}}
\def\ee{\end{equation}}
\def\beq{\begin{equation}}
\def\eeq{\end{equation}}
\def\bea{\begin{eqnarray}}
\def\eea{\end{eqnarray}}
\def\nbea{\begin{eqnarray*}}
\def\neea{\nonumber\end{eqnarray*}}
\def\bmat#1{\left(\begin{array}{#1}}
\def\emat{\end{array}\right)}
\def\bcase#1{\left\{\begin{array}{#1}}
\def\ecase{\end{array}\right.}
\def\bmini#1{\begin{minipage}{#1\textwidth}}
\def\emini{\end{minipage}}

\usepackage{wrapfig}
\graphicspath{{figures/}}
\usepackage{enumerate}

\usepackage{bbm}
\usepackage{physics}
\usepackage{graphicx}
\usepackage{dcolumn}
\usepackage{bm}
\usepackage{booktabs}
\usepackage[caption=false]{subfig}

\begin{document}
\title{Microscopic Connection between Phonon Hall Viscosity and Acoustic Faraday Effect in Spin-Orbit-Coupled Magnets}
\author{Antoine Matar}
\affiliation{Department of Physics, University of Toronto, 60 St. George St., Toronto, Ontario, Canada M5S 1A7}
\author{Haoting Xu}
\affiliation{Department of Physics, University of Toronto, 60 St. George St., Toronto, Ontario, Canada M5S 1A7}
\author{Hae-Young Kee}
\email[]{hy.kee@utoronto.ca}
\affiliation{Department of Physics, University of Toronto, 60 St. George St., Toronto, Ontario, Canada M5S 1A7}
\affiliation{Canadian Institute for Advanced Research, CIFAR Program in Quantum Materials, Toronto, Ontario, Canada M5G 1M1 }

\begin{abstract}{
The acoustic Faraday effect probes phonon Hall viscosity through the rotation of transverse acoustic polarization, conventionally interpreted in terms of circular birefringence. We show that magnetic order breaking rotational symmetry generates anisotropic phonon responses and elliptically polarized eigenmodes, so that Faraday rotation cannot be attributed to Hall viscosity alone. To elucidate their microscopic origin and consequences, we derive spin–strain couplings from a generic spin-orbit-coupled exchange model. We identify transverse strain couplings for propagation along the c-axis arising linearly from the changes in the bond-dependent interactions, along with strain-induced g-tensor contributions.  
Applying our theory to $\alpha$-RuCl$_3$, we find that static strain responses modify the extraction of Hall viscosity from Faraday rotation, yielding a value significantly different from that inferred  assuming rotational symmetry. Our results establish a microscopic connection between magnetic interaction, lattice dynamics, and phonon Hall viscosity in spin-orbit-coupled magnets.
}
\end{abstract}
\date{\today}
\maketitle

{\it Introduction} --
Hall viscosity is a dissipationless transport coefficient that characterizes the transverse response of a system with broken time-reversal symmetry and provides a sensitive probe of its underlying geometric and topological properties \cite{Mead1992RMP,Avron1995PRL,Read2009PRB,Read2011PRB,Bradlyn2012PRB,Hoyos2012PRL,Flebus2023PRL,Hoyos2014IJMPB}. In solids, Hall viscosity modifies the propagation of transverse acoustic phonons by coupling their polarization modes \cite{Barkeshli2012PRB,Shapourian2015PRB,Tuegel2017PRB,SaparovNiu2022PRB}.
In turn, this coupling can produce acoustic Faraday rotation, providing an important experimental route to infer the phonon Hall viscosity, which is otherwise difficult to access.

Spin–phonon coupling in magnetic materials can generate unconventional lattice responses, including phonon Berry curvature and transverse phonon transport. Phonon thermal Hall effects have been observed in a broad range of magnetic insulators, motivating investigations of the microscopic mechanisms underlying these geometric responses 
\cite{QinZhouShi2012PhononBerry,Takahashi2016Berry,VinklerAviv2018PRX,Ye2018PRL,Saito2019PhononBerry,Zhang2019Thermal,Ye2020PRR,Hentrich2018PRL,Li2021NatCommun,Ye2021arXiv,Kaib2021PRB,Guo2022ResonantPhonon,Lefrancois2022PRX,Zhang2024ThermalHallReview,OhNagaosa2025SpinPhononKitaev,dhakal2025theoryintrinsicphononthermal,LiYanNevidomskyy2023MagnonsPhononsRuCl3}. 
More recently, the acoustic Faraday effect has been observed in $\alpha$-RuCl$_3$, providing an experimental estimate of a large phonon Hall viscosity \cite{Ramshaw2026nature}. 
These observations highlight the importance of establishing a microscopic connection between magnetic interactions, lattice dynamics, and phonon Hall responses.

Here we develop a general microscopic theory for spin–orbit-coupled magnets and establish a connection between the acoustic Faraday effect and phonon Hall viscosity. We show that magnetic order lowers the rotational symmetry and produces elliptically, rather than circularly, polarized phonon eigenmodes, so that Faraday rotation can no longer be interpreted as a direct measure of Hall viscosity alone.
We derive an effective off-diagonal g-factor and spin-strain interactions from a generic $J$-$K$-$\Gamma$-$\Gamma^\prime$ exchange model\cite{Rau2014PRL}, and show that static elastic contributions substantially modify the inferred Hall viscosity. We also present an application of the theory to $\alpha$-RuCl$_3$ and an interpretation of the inferred Hall viscosity. 

\textit{Phonon Berry curvature and acoustic Faraday effect.}--
The adiabatic evolution of the electronic state $|\psi \rangle$ gives rise to a geometric contribution to the effective phonon dynamics. Integrating out the electronic degrees of freedom produces 
a static renormalization of the elastic constants and a geometrical contribution governed by the Berry curvature. The resulting equation of motion for the two transverse modes ${\bf u} (q) = [u_a (q), u_b (q)]$ propagating with frequency $\omega$ along the $c$-axis (${\bf q} = q {\hat c}$), in the long-wavelength limit ($q c \ll 1)$, is
\begin{equation}
\left[\rho\omega^2 {\mathbb{1}}-D(q,\omega)\right]{\bf u}(q)=0,
\end{equation}
where $c$ is the unit-cell spacing along the $c$ axis. The phonon dynamical matrix $D( q,\omega)$ is given by
\begin{equation}
D (q, \omega) = q^2
\begin{pmatrix}
C_0 - \chi_{aa} & -\chi_{ab} -i\omega\eta_{ba}\\
-\chi_{ab}+i\omega\eta_{ba} & C_0 -\chi_{bb}
\end{pmatrix},
\label{D_matrix}
\end{equation}
where $C_0$ is a bare elastic constant. The static strain response $\chi_{\mu\nu}$ is given by 
$\chi_{\mu\nu}=-\frac{1}{V}\left.\frac{\partial^2 E_0(\boldsymbol{\epsilon})} {\partial\epsilon_\mu\partial\epsilon_\nu}\right|_{\boldsymbol{\epsilon}=0}$, where $\mu,\nu = a,b$ and $E_0(\boldsymbol{\epsilon})$ denotes the instantaneous electronic ground-state energy.
The antisymmetric term $i\omega\eta_{ba}$ originates from the geometric response of the electronic state.

The geometrical  antisymmetric {\it strain} Berry curvature is defined as follows. In the long-wavelength limit, we introduce the relevant momentum-space shear strains $\epsilon_{ac} (q) = i q u_a(q)\equiv \epsilon_a(q)$ and $\epsilon_{bc} (q) = i q u_b(q) \equiv \epsilon_b (q)$.
The strain Berry curvature $\Omega_{ba}$ is determined by
\begin{equation}
\Omega_{ba} = 2 {\rm Im} \left\langle \frac{\partial \psi}{\partial \epsilon_{a}} \middle\vert  \frac{\partial \psi}{\partial \epsilon_{b}} \right\rangle,
\end{equation}
where $|\psi\rangle$ is the instantaneous electronic ground state. The phonon Hall viscosity is defined as $\eta_{ba}=\frac{\hbar}{V}\Omega_{ba}$, where $V$ is the total volume.
 In the following, $\chi_{\mu\nu}$ and $\Omega_{ba}$ denote the uniform ($q \rightarrow 0$) static strain response and strain Berry curvature, respectively.

\begin{figure}
    \centering   
     {\includegraphics[width=0.7\columnwidth]{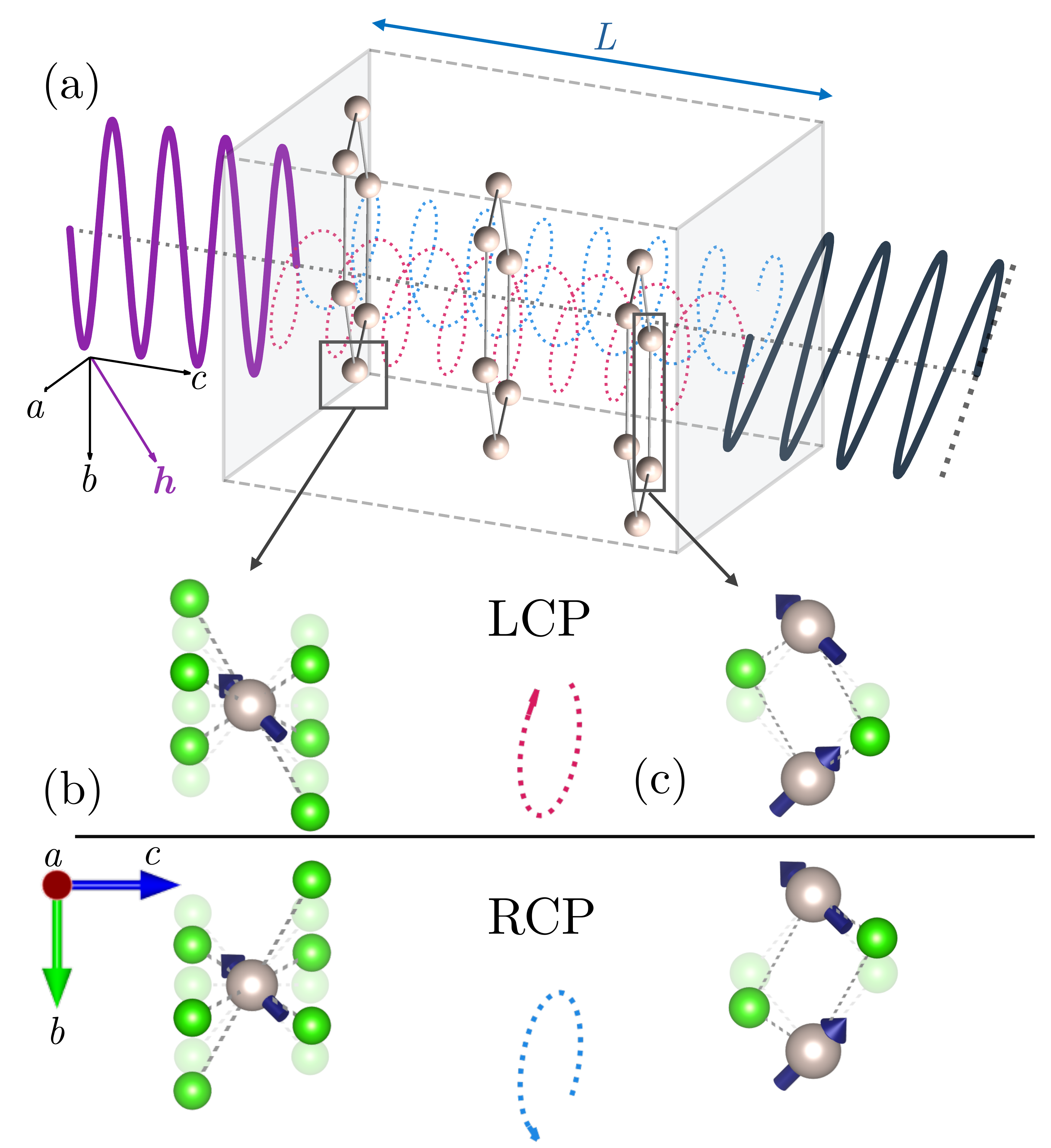}}
\caption{Schematic of the acoustic Faraday effect and the microscopic spin--phonon couplings.
(a) A transverse acoustic wave propagating along the $c$ axis, with polarization in the $ab$ plane. Hall viscosity leads to different propagation of the left- and right-circularly polarized (LCP and RCP) components, resulting in a relative phase accumulation and rotation of the transmitted polarization; the rotation is exaggerated for clarity.
The dark and light green circles indicate the ligand positions before and after the lattice distortion induced by the acoustic wave, respectively, illustrating two mechanisms of spin--phonon coupling:
(b) ligand distortions modify the local crystal field around a magnetic site M (gold circle), and hence its effective $g$-tensor, and
(c) changes in the bond angles and distances between the magnetic sites M (gold circles) and ligands (green circles) modify the exchange interaction mediated by the M--ligand--M exchange path.
}
    \label{fig:afe_hsp_fig}
\end{figure}
To characterize the acoustic Faraday response, we consider  a three-dimensional crystal with an $n$-fold rotation axis ($n>2$) along $c$, with $a$ and $b$ denoting two orthogonal directions perpendicular to the $c$ axis.
A generic experimental setup, illustrated in Fig.~\ref{fig:afe_hsp_fig}(a), consists of an incident wave with frequency $\omega$, polarized along the $b$ axis, that propagates a distance $L$ in the presence of an applied magnetic field $h$ and is then analyzed at $45^\circ$ relative to the $a$ axis in the $ab$ plane.

Conventionally, phonon Hall viscosity is inferred from the antisymmetric signal $S$ constructed from transmitted acoustic amplitudes at opposite magnetic fields. Here we show that finite static anisotropies $\chi_{\mu\nu}$ can alter both the magnitude and sign of this signal, motivating a reassessment of its interpretation as a direct measure of Hall viscosity. We further show how to determine them by combining polarization-resolved amplitude measurements with field-induced sound-velocity shifts. The antisymmetric signal $S$ is defined as 
\begin{equation}
S=
\frac{A(h)-A(-h)}
     {A (h) +A(-h)},
\label{eq:faraday_signal_definition}
\end{equation}
where $A(h)$ is the transmitted amplitude under magnetic field $h$, which is determined through the intensity of the transmitted acoustic wave for initial $i$ (for example, $b$-axis) and final $f$ ($45^{\circ}$ transmitted wave)  polarizations $I_{f \leftarrow i}$, i.e., $A = \sqrt{|I_{f\leftarrow i}|}$.

When rotational symmetry is preserved, the antisymmetric signal is proportional to the Hall viscosity, $S\propto\eta_{ba}$~\cite{Tuegel2017PRB,Ye2021arXiv,Ramshaw2026nature}.
However, when the magnetic ground state breaks the rotational symmetry, $\chi_{aa} \neq \chi_{bb}$ and $\chi_{ab} \neq 0$, making the eigenmodes elliptically polarized, and the signal can no longer be attributed to the Hall viscosity $\eta_{ba}$.
For brevity, we define $\eta \equiv \eta_{ba}$, $\chi_1 \equiv \chi_{ab}$, $\chi_3 \equiv (\chi_{aa} - \chi_{bb})/2$, $C_I \equiv C_0 - (\chi_{aa}+\chi_{bb})/2$, and $|D| \equiv \sqrt{ (\omega \eta)^2 + \chi_1^2 + \chi_3^2}$.
In the limit  $|D| \ll C_I$,
the wave-vector splitting 
between the two transverse normal modes reduces to $\delta q\approx\frac{|D| q_0}{C_I}$ where $q_0$ is the unperturbed wave vector.  The antisymmetric signal $S$ becomes
\begin{align}
S \approx {}&
-\frac{
\omega\eta
|D|\,
\sin(\delta q L)
}{
\abs{D}^2
-\chi_1\chi_3
+\chi_1\chi_3\cos(\delta q L)
}\nonumber\\
&
\times
\left\{
1+
\sqrt{
1-
\frac{
(\omega\eta)^2 |D|^2
\sin^2(\delta q L)
}{
\left[
\abs{D}^2
-\chi_1\chi_3
+\chi_1\chi_3\cos(\delta q L)
\right]^2
}
}
\right\}^{-1}.
\label{eq:signal_exact}
\end{align}
Note that when $\delta q \ll L^{-1}$, the signal reduces to the familiar relation, $S \approx -\frac{1}{2}\frac{\omega\eta}{C_I} q_0 L$; see Supplemental Material (SM) for the derivation. 

On the other hand, when  $\delta q L$ is of order 1, and taking the limit $\chi_1^2+\chi_3^2 \gg (\omega\eta)^2$, the signal $S$ becomes
\begin{equation}
S\simeq
-\frac{\omega\eta}{2\sqrt{\chi_1^2+\chi_3^2}}
\frac{
\sin\left(\delta q L\right)
}{
1-\frac{\chi_1 \chi_3}{\chi_1^2+\chi_3^2}
\left[1-\cos\left(\delta q L\right)\right]
}.
\label{eq:faraday_anisotropic}
\end{equation}
Note that the signal can change sign, as observed, for example, in $\alpha$-RuCl$_3$\cite{Ramshaw2026nature}, implying that $\chi_1$ and $\chi_3$ are significant, and thus should not, in general, be neglected in the analysis.
The inferred Hall viscosity $\eta_0$, obtained assuming $\chi_1 = \chi_3 =0$, is related to the actual viscosity $\eta$ as follows: 
\begin{equation}
\frac{\eta_0}{\eta}
= 
\frac{\sin\left(\delta q L\right)/(\delta q L)}{1-\frac{\chi_1 \chi_3}{\chi_1^2 + \chi_3^2} (1-\cos\delta q L)}.
\label{eq:eta_inferred}
\end{equation}
In particular, when $\delta q L = \pi$, $\eta_0 =0$, while the true value of $\eta$ is finite. Thus, including the static anisotropy is crucial when extracting the Hall viscosity from the measured signal $S$.
 
\begin{figure}
    \centering
    \includegraphics[width=1.0\linewidth]{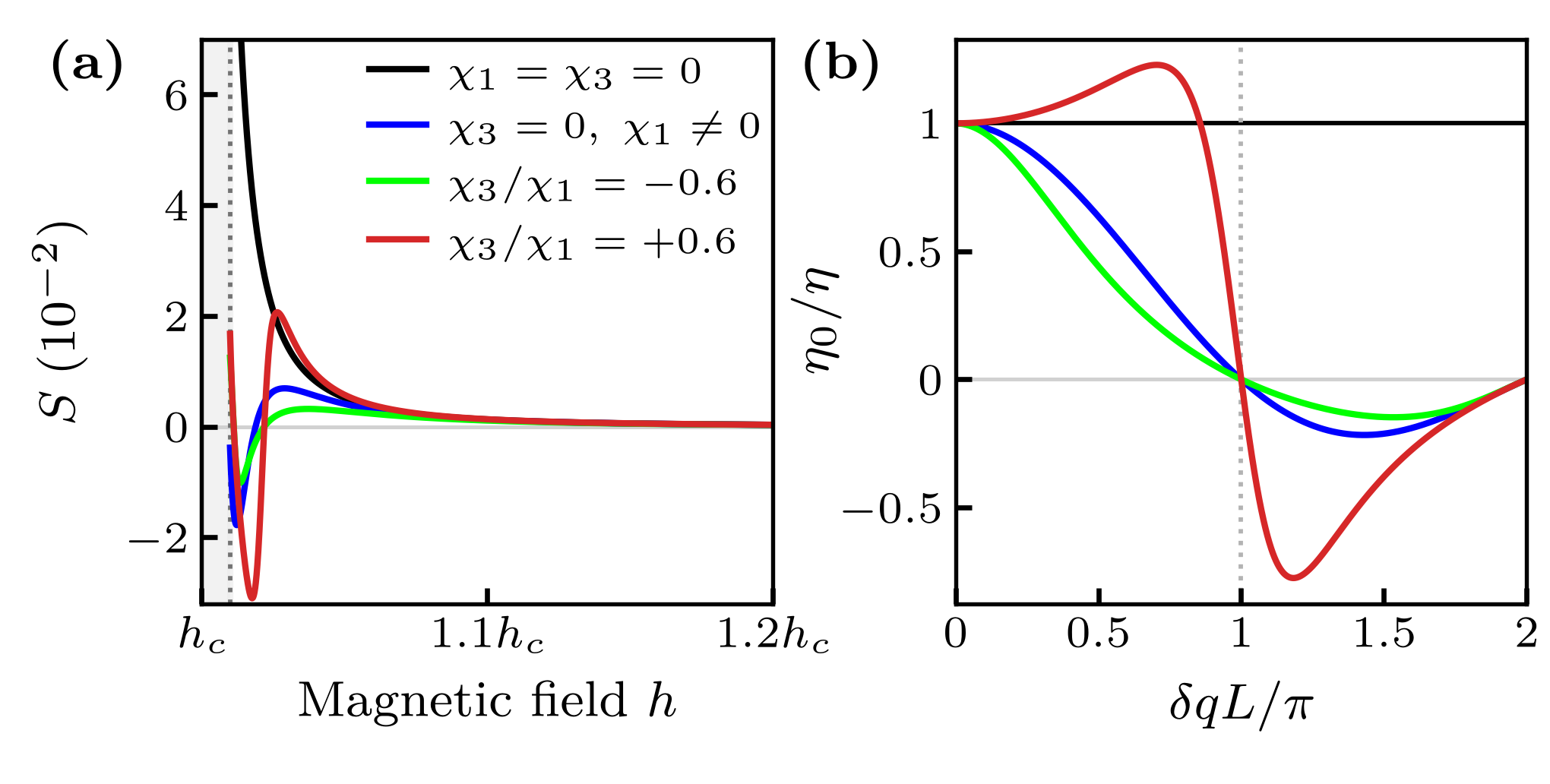}
\caption{Phenomenological model to illustrate the effects of static spin-induced anisotropy $\chi_3/\chi_1$ on the acoustic Faraday response. (a) Antisymmetric signal $S$ obtained by numerically propagating the eigenmodes of Eq.~\eqref{D_matrix}. We show an example of a field-driven transition at $h_c$ to a spin-polarized phase. 
For $h>h_c$, we assume $\Delta(h)\propto h-h_c$, $\chi_1,\chi_3\propto\Delta^{-1}$, and $\eta\propto\Delta^{-2}$.
Blue, green, and red correspond to $\chi_3/\chi_1=0,-0.6,+0.6$, respectively, with fixed $\chi_1(h)$ and $\eta(h)$. The black curve assumes rotational symmetry ($\chi_1=\chi_3=0$). 
The gray region indicates  where the perturbative approach breaks down. (b) Ratio of the Hall viscosity $\eta_0/\eta$ 
versus
$\delta qL/\pi$, with the same color convention as (a).}
\label{fig:generic_faraday}
\end{figure}

To illustrate the importance of broken rotational symmetry, we consider a simple model  and study $S$ as a function of the field $h$ and $\eta_0/\eta$ as a function of $\delta q L$ for different values of $\chi_3/\chi_1$ in Fig.~\ref{fig:generic_faraday}. 
As an example, we demonstrate a field-driven transition at $h_c$ into a spin-polarized
phase, where the spin excitation gap opens linearly as $\Delta(h)\propto(h-h_c)$ for $h>h_c$. Assuming the response is dominated by this low-energy excitation,
perturbation theory in the spin-phonon coupling motivates
$\chi_1,\chi_3\sim\Delta(h)^{-1}$ and $\eta\sim\Delta(h)^{-2}$. We choose the overall response scales to be representative of those obtained from the microscopic calculation below. 
As shown in Fig.~\ref{fig:generic_faraday}(a), even the smooth field dependence can generate oscillations in $S$ (colored lines) near $h_c$ due to the accumulated phase $\delta q L$ when either $\chi_1$ or $\chi_3$ is finite, in contrast to the rotationally invariant case (black line).
Similarly, $\eta_0/\eta$ is unity for the rotationally invariant system by definition (black line) in Fig.~\ref{fig:generic_faraday}(b), while finite static responses induce a sign change when the phase $\delta q L$ crosses $\pi$.

To extract phonon Hall viscosity from the acoustic Faraday signal, one requires accounting for the static anisotropies $\chi_1$ and $\chi_3$. We propose that these quantities can be measured through the polarization-resolved sound velocity measurements. 
First, the ratio of the off-diagonal response $\chi_1$ to the diagonal anisotropy $\chi_3$ can be determined by combining intensity measurements for different incident polarizations $i$ and transmitted polarizations $f$, $I_{f\leftarrow i}=A_{f\leftarrow i}^2$. 
Using the linear polarization of $\pm \pi/4$ ($\pm 45^\circ$) relative to the $a$ axis in the $ab$ plane, one obtains, for $\chi_3\neq0$ (see SM for details),
\begin{equation}
\frac{\chi_1}{\chi_3}
=
\tan\left[
\frac{1}{2}
\operatorname{tan^{-1}}\left(\frac{I_{a\leftarrow \pi/4}-I_{-\pi/4\leftarrow a}}{I_{-\pi/4\leftarrow \pi/4}-I_{b\leftarrow a}}\right)
\right].
\label{eq:chi_inferred}
\end{equation}
Having determined $\chi_1/\chi_3$, we obtain $\chi_3$ from polarization-resolved sound-velocity shifts. Let $v_a(h)$ and $v_b(h)$ denote the effective phase
velocities extracted from the $a\rightarrow a$ and
$b\rightarrow b$ transmission channels, respectively.
Their difference can be determined from calibrated
polarization-resolved ultrasonic phase measurements.
For $|D|\ll C_I$, we obtain
\begin{equation}
\frac{\rho\omega L}{2C_0}(v_b(h)-v_a(h))
\simeq
\tan^{-1}\left[
\frac{\chi_3}{|D|}
\tan\left(\frac{\delta qL}{2}\right)
\right].
\label{eq:velocity_difference}
\end{equation}
Taking the limit $(\omega\eta)^2\ll\chi_1^2+\chi_3^2$
and $C_I\simeq C_0$, we obtain
\begin{equation}
\begin{aligned}
\chi_3\simeq{}&
\frac{2C_0}
{q_0L\sqrt{1+(\frac{\chi_1}{\chi_3})^2}}
\\
&\times\tan^{-1}\left[
\sqrt{1+\left(\frac{\chi_1}{\chi_3}\right)^2}
\tan\left(
\frac{\rho\omega L}{2C_0}(v_b(h)-v_a(h)
\right)
\right].
\end{aligned}
\label{eq:chi3_inferred}
\end{equation}
Together with $\chi_1/\chi_3$, this determines both
static responses entering the Faraday signal.

\textit{Spin-phonon interaction and resultant Berry curvature.}--- To apply this framework microscopically, we next derive the spin-strain interactions starting from local pseudospins and the generic $J$-$K$-$\Gamma$-$\Gamma^\prime$ model \cite{Rau2014PRL} with interlayer coupling for spin-orbit-coupled magnets. The first contribution to spin-strain coupling, denoted by ${\cal H}^{(1)}_{\text {int}}$, arises from the change of the ligand environment surrounding the magnetic ion.
The single-site off-diagonal g-factor via the spin--strain interaction takes the following form, 
\begin{equation}
    {\cal H}_{\mathrm{int}}^{(1)}=-\sum_i {\mu_B}\tilde g\epsilon_{a}(i)(s^a_i h_c+s^c_i h_a)+(a\rightarrow b), 
    \label{effective_g}
\end{equation}
where $s^\alpha$ $(\alpha=a,b,c)$ denotes the spin-$1/2$ pseudospin operator along the global crystallographic axes and the external field ${\bf h} = (h_a,h_b, h_c)$. Note that we define the strain change $\epsilon_a (i)$ locally at each site $i$, since our microscopic model is formulated in the real space. To obtain the static and dynamic responses discussed below, we Fourier transform the Hamiltonian and take the long-wavelength limit. 

The second contribution to spin-strain coupling, $H_{\mathrm{int}}^{(2)}$, arises from the change of the hopping integrals. These hopping modulations alter the spin-exchange interactions and thereby generate an additional spin-strain coupling. 
We consider a three-dimensional R${\bar 3}$ lattice including the dominant Kitaev interlayer coupling \cite{Rau2014PRL,RauKee2014Trigonal,Cen2025PRB}. In the crystallographic $a$-$b$-$c$ coordinate system, the $J$-$K$-$\Gamma$-$\Gamma'$ Hamiltonian, ${\cal H}^0_{\mathrm{spin}}=\sum_{\langle ij\rangle\in\gamma}{\cal H}_{ij}^{\gamma}$, is written as\cite{Chaloupka2015PRB,Rousochatzakis2024ROPP,Matsuda2025RMP,Cen2025PRB}, 
\begin{equation}
\renewcommand{\arraystretch}{1.5}
\begin{array}{l}
{\cal H}_{ij}^\gamma\!=\!J_{ab}(s_{i}^{a}s_{j}^{a}+s_{i}^{b}s_{j}^{b})+J_{c}s_{i}^{c}s_{j}^{c}\\~~
+J_{\pm\pm}\left[{\rm c}_\gamma(s_{i}^{a}s_{j}^{a}-s_{i}^{b}s_{j}^{b})-{\rm s}_\gamma(s_{i}^{a}s_{j}^{b}+s_{i}^{b}s_{j}^{a})\right]\\~~
-\sqrt{2} J_{\pm c}\left[{\rm c}_\gamma(s_{i}^{a}s_{j}^{c}+s_{i}^{c}s_{j}^{a})+{\rm s}_\gamma(s_{i}^{b}s_{j}^{c}+s_{i}^{c}s_{j}^{b})\right] + {\cal H}_{\rm layer},
\end{array}
\label{eq:Hamiltonian2}
\end{equation}
where ${\rm c}_\gamma\!\equiv\!\cos\phi_\gamma$, ${\rm s}_\gamma\!\equiv\!\sin\phi_\gamma$, $\phi_{\gamma}\!=\!0$, $\frac{2\pi}{3}$, and $\frac{4\pi}{3}$ for $\gamma\!=\!z$-, $x$-, and $y$-bond, respectively, and 
$J_{\pm\pm}= \frac{1}{3}K+\frac{2}{3}(\Gamma - \Gamma'), \;\; J_{\pm c}=\frac{1}{3}K-\frac{1}{3} (\Gamma - \Gamma'), \;\;
J_{ab} =J+ J_{\pm c} - \Gamma',\;\; J_{c}=J+ J_{\pm\pm} + 2 \Gamma'\,.$ 
${\cal H}_{\rm layer}$ is the Kitaev interaction $K_c$ between the layers for R${\bar 3}$ structure, responsible for the zigzag ordering pattern along the $c$-axis~\cite{Cen2025PRB}.
\begin{figure}[t]
    \centering
    \includegraphics[width=1.0\columnwidth]{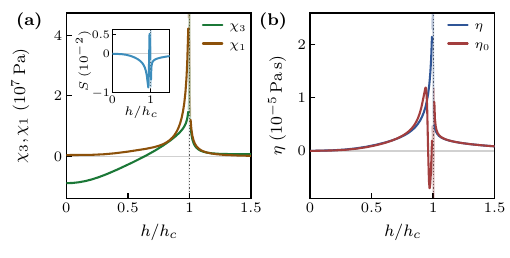}
    \caption{Acoustic Faraday response obtained using the $J$--$K$--$\Gamma$--$\Gamma'$-$K_c$ model with $J=0$, $K=-2~\mathrm{meV}$, $\Gamma=1~\mathrm{meV}$, $\Gamma'=-0.25~\mathrm{meV}$, and $K_c =0.005 \mathrm{meV}$, averaged equally over the $X$- and $Y$-zigzag domains. The magnetic field lies in the $bc$-plane at an angle $\theta=55^\circ$ from the $c$-axis. See the SM for detailed parameter values. (a) Static anisotropic responses, $\chi_3$ and $\chi_1$ as functions of field $h$. The inset shows the corresponding antisymmetric Faraday signal \(S\). The vertical dotted line marks $h_c$, and
    the region around $h_c$ is omitted where the perturbation theory breaks down. (b) Hall viscosity $\eta$ and the inferred value $\eta_0$ obtained using the rotationally symmetric Faraday expression, with a significant difference between them near $h_c$. 
}
    \label{fig:microscopic_faraday}
\end{figure}

While the strain affects all hopping processes, symmetry identifies $J_{\pm c}$ as the leading exchange channel contributing to $\Omega_{\epsilon_{b},\epsilon_{a}}$. 
Note that $J_{\pm c}$ contains the $K$ and $\Gamma$ interactions (and a smaller $\Gamma'$ term), signaling their significant contributions to the Faraday signal. 
The strain-induced modulation of the hopping amplitudes generates the following spin-phonon coupling (see SM for the detailed derivation): 
\begin{align}
    {\cal H}_{\mathrm{int}}^{(2)} &= \sum_{\langle ij \rangle_\parallel \in \gamma} \epsilon_{a} (i)
    \Big\{ (g_1 {\rm c}^2_\gamma  + g_2 {\rm s}^2_\gamma)(s^a_i s^c_j + s^c_{i}s^a_{j}) \nonumber\\
    & + {\rm s}_\gamma {\rm c}_\gamma (g_1 - g_2)(s^b_{i}s^c_{j} + s^c_{i}s^b_j) \Big\} 
    + \begin{pmatrix} a \leftrightarrow b \\ {\rm c}_\gamma \leftrightarrow {\rm s}_\gamma \end{pmatrix}
\end{align}
where the magnetoelastic couplings $g_1 \equiv -\sqrt{2}\frac{\partial J_{\pm c}}{\partial \epsilon_{a}} \eval_{z-bond}$ and $g_2 \equiv-\sqrt{2} J_{\pm c}\frac{\partial \phi_\gamma}{\partial \epsilon_{b}}\eval_{z-bond}$.
Here we ignore the interlayer coupling correction as the out-of-plane coupling is much smaller in van der Waals materials, while keeping the unstrained layer coupling in ${\cal H}_{\rm spin}^0$.

With the spin-strain interaction ${\cal H}_{\mathrm{int}} = {\cal H}_{\mathrm{int}}^{(1)} + {\cal H}_{\mathrm{int}}^{(2)}$, we evaluate the static and geometric responses within the Born-Oppenheimer approximation, 
\begin{eqnarray}
    \chi_{\mu\nu}
   & =&
    \frac{2}{V}\,\mathrm{Re}
    \sum_{m\neq0}
    \frac{
    \langle0|\partial_{\epsilon_\mu} {\cal H}_{\mathrm{int}} |m\rangle
    \langle m|\partial_{\epsilon_\nu} {\cal H}_{\mathrm{int}}|0\rangle
    }{E_m-E_0},\nonumber\\
    \Omega_{ba}
   & =&
    2\,\mathrm{Im}
    \sum_{m\neq0}
    \frac{
    \langle0|\partial_{\epsilon_a} {\cal H}_{\rm int} |m\rangle
    \langle m|\partial_{\epsilon_b} {\cal H}_{\rm int} |0\rangle
    }{(E_m-E_0)^2}.
    \label{eq:Berry_curvature}
\end{eqnarray}
where $|m\rangle$ and $|0\rangle$ are the eigenstates of the unstrained spin Hamiltonian ${\cal H}_{\mathrm{spin}}^0$. 

{\it Application to $\alpha$-RuCl$_3$: Estimating the off-diagonal g-tensor and magnetoelastic couplings.} ---
We now combine the general framework for the acoustic Faraday signal with the microscopic model and apply it to 
$\alpha$-RuCl$_3$, which exhibits zigzag magnetic order \cite{Rau2014PRL,RauKee2014Trigonal,Johnson2015MonoclinicStructure,Baek2017FieldInducedQSL,Zheng2017FieldInducedQSL,Wolter2017prb,Banerjee2018FieldINS,Gordon2019FieldRevealedKSL,Jiang2019FieldInducedQSLKH,Balz2021InterlayerIntermediateOrder,Zhou2023HighFieldPhase,Cen2025PRB}. To connect the spin--phonon couplings introduced above
to microscopic parameters, we estimate $\tilde g$, $g_1$, and $g_2$ from
real-space integrals evaluated using multi-zeta Clementi--Roetti atomic
orbitals (see SM for details). 
We first estimate ${\tilde g}$ in Eq. \eqref{effective_g}. This requires the orbital--strain coupling strength by evaluating the change in
the local octahedral crystal potential under shear strain and projecting it
onto the corresponding quadrupolar operators. We find $\tilde g\simeq6.2$  (see SM for the derivation). The magnetoelastic couplings $g_1$ and $g_2$ arise from the modulation of the hopping amplitude of  Ru-Cl-Ru pathways. They are evaluated by perturbing the positions of the intermediate Cl ions and computing the orbital overlap integrals. The resulting estimates depend on the orbital energies, and ligand geometry. We find $g_1 \sim10  J_{\pm c}$, while $g_2$ is small, $O(1) J_{\pm c}$, and has a sign sensitive to the microscopic parameters (see SM). In Fig.~3, we use $g_2 = -2.6 J_{\pm c}$, which yields a signal $S$ closest to that reported in Ref.~\cite{Ramshaw2026nature}.

To evaluate the resulting acoustic Faraday response, we model the magnetic sites of $\alpha$-RuCl$_3$ as stacked honeycomb layers with
R$\bar{3}$ crystal symmetry, which preserves a threefold rotation axis
along $c$. The magnetic interactions are described by a
$J$-$K$-$\Gamma$-$\Gamma'$-$K_c$ model with a zigzag-ordered ground state.
Within linear spin-wave theory, obtained through the Holstein--Primakoff transformation, the low-energy excited states are magnons. The matrix elements entering Eq.~\eqref{eq:Berry_curvature} contain contributions from one- and
two-magnon states. We compute the Hall viscosity, static strain susceptibility, and the resulting antisymmetric signal, as depicted in Fig.~\ref{fig:microscopic_faraday}. 

The signal $S$ shown in the inset of Fig.~\ref{fig:microscopic_faraday}(a) peaks at values of order $10^{-2}$ before the magnetic transition, strongly oscillates around $h_c$, and then decays with $h$ in the polarized state. This behavior is consistent with the measurement in Ref.~\cite{Ramshaw2026nature}, indicating significant $\chi_1$ and $\chi_3$. We find $\chi_1$ is indeed large near the transition as shown in Fig.\ref{fig:microscopic_faraday}(a). The  Hall viscosity $\eta$ estimated from our theory peaks at approximately $2\times10^{-5}\,\mathrm{Pa\cdot s}$, which is around twice the value $\eta_0$ that would be inferred from the same signal by assuming rotational symmetry, as shown in Fig.~\ref{fig:microscopic_faraday}(b). 
Our results highlight the importance of broken rotational symmetry in relating the acoustic Faraday response to phonon Hall viscosity. 

{\it Summary and Discussion.} -- 
We establish a theoretical framework relating spin-strain interactions and the acoustic Faraday effect. We note that broken rotational symmetry permits inequivalent diagonal and additional off-diagonal components of the phonon dynamical matrix, resulting in elliptically polarized eigenmodes. These contributions modify the polarization rotation, which cannot be fully described by Hall-viscosity-induced circular birefringence alone. We further propose polarization-resolved acoustic measurements to independently determine the static anisotropic responses, enabling a more reliable extraction of Hall viscosity.

We derive the microscopic spin--strain interactions from the generic $J$-$K$-$\Gamma$-$\Gamma^\prime$ exchange model  and the field-induced g-tensor in spin-orbit-coupled magnets. We apply them to $\alpha$-RuCl$_3$, considering two zigzag domains under an applied magnetic field. Our calculations reproduce oscillatory features in the antisymmetric signal $S$, suggesting significant contributions from often-neglected static phonon responses. Incorporating these contributions, the phonon Hall viscosity is approximately twice the value obtained assuming rotational symmetry. 
 Our finding of a larger Hall viscosity may have implications for the spin-liquid interpretation of the intermediate-field phase\cite{Kasahara_2018}, particularly in light of previous theoretical studies suggesting that a field-induced quantum-disordered phase emerges only for fields near the c-axis direction \cite{Gordon2019FieldRevealedKSL,Lee2020}.
A microscopic investigation of phonon Hall viscosity in field-induced quantum-disordered states of extended Kitaev models remains an open direction for future research.  Our approach extends naturally to  other spin--orbit-coupled magnets beyond $\alpha$-RuCl$_3$. 

\textit{Acknowledgment-}
We thank B. Ramshaw for useful discussion. This work is supported by the NSERC Discovery Grant No. 2022-04601 and NSERC CREATE program No. 575280-2023. H. Y. K. acknowledges support from the Canada Research Chairs Program No. CRC-2019-00147.
This research was enabled in part by support provided by Compute Ontario,  Calcul Québec, and the Digital Research Alliance of Canada.

\bibliographystyle{apsrev4-2}
\bibliography{cites}
\end{document}

\title{Supplementary Material "Microscopic Connection between Phonon Hall Viscosity and Acoustic Faraday Effect in Spin-Orbit-Coupled Magnets"}
\author{Antoine Matar}
\affiliation{Department of Physics, University of Toronto, 60 St. George St., Toronto, Ontario, Canada M5S 1A7}
\author{Haoting Xu}
\affiliation{Department of Physics, University of Toronto, 60 St. George St., Toronto, Ontario, Canada M5S 1A7}
\author{Hae-Young Kee}
\email[]{hy.kee@utoronto.ca}
\affiliation{Department of Physics, University of Toronto, 60 St. George St., Toronto, Ontario, Canada M5S 1A7}
\affiliation{Canadian Institute for Advanced Research, CIFAR Program in Quantum Materials, Toronto, Ontario, Canada M5G 1M1 }
\date{\today}

\maketitle

\section{Phonon equation of motion and acoustic Faraday effect}
In this section, we derive the effective phonon dynamics induced by spin--strain coupling and examine how elastic anisotropy modifies the acoustic Faraday signal used to probe phonon Hall viscosity. Within the Born--Oppenheimer approximation, the
electronic degrees of freedom, represented in our
framework by the spin sector, evolve adiabatically
with the atomic configuration. We describe the
acoustic deformation by a continuous displacement
field $\mathbf u(\mathbf r,t)$, where $\mathbf r$
denotes the spatial position, and the corresponding
strain tensor $\boldsymbol\epsilon(\mathbf r,t)$.
For transverse waves propagating along the $c$ axis,
the relevant shear strains are
\begin{align}
\epsilon_a(\mathbf r,t)
&\equiv\epsilon_{ca}(\mathbf r,t)
=\partial_c u_a+\partial_a u_c,\\
\epsilon_b(\mathbf r,t)
&\equiv\epsilon_{bc}(\mathbf r,t)
=\partial_c u_b+\partial_b u_c.
\end{align}
Following the notation of the main text, we use
$\mu,\nu\in\{a,b\}$ to label these two strain
components. We denote the equilibrium unit-cell positions by
$\mathbf R$ and the position of sublattice $l$
within a unit cell by $\mathbf r_l$. The position
of magnetic site $i$ is then
$\mathbf r_i=\mathbf R+\mathbf r_l$.

The spin ground state perturbed by the
spin--phonon interaction ${\cal H}_{\rm int}$
can then be expressed to first order as
\begin{equation}
    |\psi\rangle
    =
    |0\rangle
    +
    \sum_{m\neq 0}
    \frac{\langle m|{\cal H}_{\rm int}|0\rangle}
    {E_0-E_m}
    |m\rangle,
    \label{eq:spin_gs}
\end{equation}
where $|0\rangle$ and $|m\rangle$ are the ground and excited states of the unperturbed spin Hamiltonian, with corresponding energies $E_0$ and $E_m$. To linear order in strain, the spin--phonon interaction can be decomposed as $   {\cal H}_{\text{int}} = \sum_{\mu} \frac{\partial {\cal H}_{\text{int}}}{\partial \epsilon_\mu} \epsilon_\mu.$
    
To study the phonon dynamics, we integrate out the spin degrees
of freedom using the time-dependent variational
principle~\cite{Mead1992RMP,SaparovNiu2022PRB}:
\begin{equation}
    \mathcal{L}_{\text{ph}}
    = \frac{V}{N}\sum_{\mathbf{R}} \left\{
    \frac{\rho}{2}\dot{\mathbf{u}}^2(\mathbf{R})
    - V_{\text{ph}}(\mathbf{R})
    \right\}
    - \left(
    \langle \psi | {\cal H}_{\rm spin} | \psi \rangle
    - E_0
    \right)
    + \langle \psi | i\hbar\partial_t | \psi \rangle,
    \label{eq:variational_principle_lagrangian}
\end{equation}
where ${\cal H}_{\rm spin}={\cal H}_{\rm spin}^{0}
+{\cal H}_{\rm int}$, $V$ is the total volume,
and $N$ is the number of unit cells.
The bare elastic phonon energy is expressed using the
Voigt stiffness matrix $C_{\mu\nu}$ as
\begin{equation}
    V_{\text{ph}}(\mathbf{R})
    = \frac{1}{2}\sum_{\mu\nu}
    C_{\mu\nu}\epsilon_\mu(\mathbf{R})
    \epsilon_\nu(\mathbf{R}).
\end{equation}

The spin ground-state energy provides a static correction
to the elastic energy. Expanding to second order in strain
and omitting the linear term about equilibrium, we obtain
\begin{equation}
\begin{aligned}
    \langle \psi | {\cal H}_{\rm spin} | \psi \rangle-E_0
    &\simeq
    -\sum_{m\neq 0}
    \frac{
    |\langle m|{\cal H}_{\rm int}|0\rangle|^2
    }{E_m-E_0}\\
    &\equiv
    -\frac{V}{2}\sum_{\mu\nu}
    \chi_{\mu\nu}\epsilon_\mu\epsilon_\nu,
\end{aligned}
\end{equation}
where $\chi_{\mu\nu}$ is the symmetric static strain
susceptibility that renormalizes the stiffness tensor.
Here $V$ is the total volume.
Assuming the spin ground state does not depend on time explicitly, we can write the rightmost term in Eq.~\eqref{eq:variational_principle_lagrangian} using the strain-space Berry connection $A_\mu$: 
\begin{equation}
    \langle\psi | i\partial_t | \psi \rangle = \sum_{\mu} \langle \psi | i\frac{\partial}{\partial\epsilon_\mu} | \psi \rangle \dot{\epsilon}_\mu = \sum_{\mu} A_\mu \dot{\epsilon}_\mu,
\end{equation}
and the corresponding Berry curvature
\begin{equation}
\begin{aligned}
    \Omega_{ba}
    &=\frac{\partial A_a}{\partial\epsilon_b}
    -\frac{\partial A_b}{\partial\epsilon_a}=2\,\mathrm{Im}
    \left\langle
    \frac{\partial\psi}{\partial\epsilon_a}
    \middle|
    \frac{\partial\psi}{\partial\epsilon_b}
    \right\rangle.
\end{aligned}
\end{equation}
As in the main text, the Hall viscosity is
$\eta_{ba}=\hbar\Omega_{ba}/V$.

Substituting a plane-wave ansatz $\mathbf{u}(\mathbf{R}, t) = \mathbf{u}_0 e^{i(\mathbf{q} \cdot\mathbf{R} - \omega t)}$ for the two transverse acoustic modes propagating along the $c$-axis, the Euler-Lagrange equation of motion becomes Eq.~(1) in the main text:
\begin{equation}
    \rho \omega^2 \begin{pmatrix}
        u_a \\ u_b
    \end{pmatrix} = q^2 \begin{pmatrix}
        C_0 -\chi_{aa}& -\chi_{ab}-i\omega\eta_{ba} \\ -\chi_{ab}+i\omega\eta_{ba} &  C_{0}-\chi_{bb}
    \end{pmatrix} \begin{pmatrix}
        u_a \\ u_b
    \end{pmatrix}.
    \label{eq:Equation_of_motion_2}
\end{equation}
The above $2\times2$ matrix can be expressed in terms of
Pauli matrices as
\begin{equation}
C_I\mathbb{1}-\chi_3\sigma_z
-\chi_1\sigma_x+\omega\eta\sigma_y,
\end{equation}
where $C_I=C_0-(\chi_{aa}+\chi_{bb})/2$,
$\chi_3=(\chi_{aa}-\chi_{bb})/2$,
$\chi_1=\chi_{ab}$, $\eta=\eta_{ba}$. We also introduce 
$|D|=\sqrt{\chi_1^2+\chi_3^2+(\omega\eta)^2}$,
as in the main text. The corresponding eigenvalues are
\begin{equation}
\lambda_\pm=C_I\pm|D|,
\end{equation}
with normalized eigenvectors $\mathbf u_\pm$.
In general, finite static anisotropy produces elliptically
rather than circularly polarized eigenmodes.

The dispersion relations are
\begin{equation}
\rho\omega^2=q_\pm^2(C_I\pm|D|),
\end{equation}
giving
\begin{equation}
q_\pm=\omega\sqrt{\frac{\rho}{C_I\pm|D|}}
\simeq\omega\sqrt{\frac{\rho}{C_I}}
\left(1\mp\frac{|D|}{2C_I}\right),
\end{equation}
where the expansion assumes $|D|\ll C_I$.

Consider an incident wave polarized along the $b$-axis, $\mathbf u_i=(0,1)^T$, propagating a distance $L$ before being detected at $45^\circ$ relative to the $a$ axis, $\mathbf u_f=(1,1)^T/\sqrt{2}$.
The transmitted wave is a superposition of the two
normal modes, with contributions
\begin{equation}
A_\pm=
(\mathbf u_\pm^\dagger\mathbf u_i)
(\mathbf u_f^\dagger\mathbf u_\pm)e^{iq_\pm L}
=\mathbf u_f^\dagger
(\mathbf u_\pm\mathbf u_\pm^\dagger)\mathbf u_ie^{iq_\pm L}.
\end{equation}
The measured amplitude is $A=|A_++A_-|$.

Using the eigenmode projectors
\begin{equation}
\mathbf u_\pm\mathbf u_\pm^\dagger
=\frac12\left[
\mathbb{1}\mp
\frac{\chi_1\sigma_x+\chi_3\sigma_z
-\omega\eta\sigma_y}{|D|}
\right],
\end{equation}
the sum of the two contributions becomes
\begin{align}
A_++A_-={}&
e^{i(q_++q_-)L/2}
\mathbf u_f^\dagger
\left[
\cos\left(\frac{\delta qL}{2}\right)\mathbb{1}
\right.\left.
+i\frac{\sin(\delta qL/2)}{|D|}
(\chi_1\sigma_x+\chi_3\sigma_z
-\omega\eta\sigma_y)
\right]\mathbf u_i,
\end{align}
where $\delta q=q_--q_+$.
Substituting the incident and detected polarizations gives
\begin{align}
A_++A_-={}&
\frac{e^{i(q_++q_-)L/2}}{\sqrt{2}}
\left[
\cos\left(\frac{\delta qL}{2}\right)
-\frac{\omega\eta}{|D|}
\sin\left(\frac{\delta qL}{2}\right)
+i\frac{\chi_1-\chi_3}{|D|}
\sin\left(\frac{\delta qL}{2}\right)
\right].
\end{align}
Taking the magnitude and assuming that $\chi_1$ and
$\chi_3$ are even under field reversal while $\eta$
is odd, we obtain
\begin{align}
A(\pm h)=\frac{1}{\sqrt{2}}\Bigg[
&1-\frac{\chi_1\chi_3}{|D|^2}
\left(1-\cos(\delta qL)\right)
\mp\frac{\omega\eta}{|D|}
\sin(\delta qL)
\Bigg]^{1/2}.
\end{align}

The antisymmetric Faraday signal is defined as
\begin{equation}
S=\frac{A(h)-A(-h)}{A(h)+A(-h)}.
\end{equation}
For an incident and analyzer polarizations
specified above, the signal is given by
\begin{align}
S={}&
-\frac{
\omega\eta|D|\sin(\delta qL)
}{
|D|^2-\chi_1\chi_3
+\chi_1\chi_3\cos(\delta qL)
}
\left[
1+
\sqrt{
1-
\frac{
(\omega\eta)^2|D|^2\sin^2(\delta qL)
}{
\left[
|D|^2-\chi_1\chi_3
+\chi_1\chi_3\cos(\delta qL)
\right]^2
}
}
\right]^{-1}.
\label{eq:signal_exact}
\end{align}
For $|D|\ll C_I$, the splitting reduces to
$\delta q\simeq q_0|D|/C_I$, where
$q_0=\omega\sqrt{\rho/C_0}$ and $C_I\simeq C_0$.
The signal depends on the incident and detected
polarizations and vanishes when they are identical.
Moreover, only odd powers of $\eta$ contribute to $S$.

In the limit $\delta qL\ll1$, without any restriction
on the relative magnitudes of $\chi_1^2+\chi_3^2$ and $(\omega\eta)^2$ the signal reduces to
\begin{equation}
S\simeq-\frac12\frac{\omega\eta}{C_I}q_0L.
\label{eq:signal_smalltheta}
\end{equation}
On the other hand, when $\delta qL$ is of order $1$
and $C_I^2\gg\chi_1^2+\chi_3^2\gg(\omega\eta)^2$, the
wave-vector splitting becomes
$\delta q\simeq q_0\sqrt{\chi_1^2+\chi_3^2}/C_I$,
and the signal simplifies to
\begin{equation}
S\simeq
-\frac{\omega\eta}{2\sqrt{\chi_1^2+\chi_3^2}}
\frac{
\sin(\delta qL)
}{
1-\frac{\chi_1\chi_3}{\chi_1^2+\chi_3^2}
\left[1-\cos(\delta qL)\right]
}.
\label{eq:SM:signal}
\end{equation}
This expression shows how the static anisotropies
$\chi_1$ and $\chi_3$ modify the Faraday signal through
the accumulated phase $\delta qL$.

For the numerical results in Fig.~3 of the main text,
we use an acoustic frequency $\omega/2\pi=1.5~\mathrm{GHz}$,
propagation length $L=210~\mu\mathrm{m}$,
bare elastic constant $C_0\simeq8\times10^9~\mathrm{Pa}$, and mass density $\rho=3000~\mathrm{kg/m^3}$.
\section{Estimating $\chi_3$ and $\chi_1$ from polarization-resolved ultrasound}
In this section, we show how the elastic anisotropies $\chi_3$ and $\chi_1$ can be extracted from polarization-resolved ultrasound measurements. We first obtain the ratio $\chi_1/\chi_3$ from transmission intensities and then combine it with the phase-velocity shifts in the $a\rightarrow a$ and $b\rightarrow b$ channels to determine their absolute magnitudes. Let $I_{f\leftarrow i}=|A_{f\leftarrow i}|^2$
denote the normalized transmitted intensity for incident polarization $i$ and detected polarization $f$. We find:
\begin{align}
&
I_{-\pi/4\leftarrow a}-I_{a\leftarrow \pi/4} =
\frac{-2\chi_1\chi_3}
{\chi_3^2+\chi_1^2+(\omega\eta)^2}
\sin^2\left(\frac{\delta qL}{2}\right),
\\
&
I_{-\pi/4\leftarrow \pi/4}-I_{b\leftarrow a}
\nonumber=
\frac{\chi_3^2-\chi_1^2}
{\chi_3^2+\chi_1^2+(\omega\eta)^2}
\sin^2\left(\frac{\delta qL}{2}\right),
\end{align}
where $\pi/4$ refers a polarization tilted by $\pi/4$ from $a$ in the $ab$ plane. The dependence on the accumulated phase and Hall viscosity is common to both terms and therefore cancels when their relative angle is taken. For $\chi_3\neq0$,
\begin{equation}
\frac{\chi_1}{\chi_3}
=
\tan\left[
\frac{1}{2}
\operatorname{atan2}\left(I_{a\leftarrow \pi/4}-I_{-\pi/4\leftarrow a},I_{-\pi/4\leftarrow \pi/4}-I_{b\leftarrow a}
\right)\right].
\label{eq:chi_delta_SM}
\end{equation}
This expression is exact and does not require $\delta qL\ll1$. It becomes ill-defined when $\sin(\delta qL/2)=0$, for which both terms vanish and the polarization-resolved intensities contain no information about the ratio.

We next use the difference between the phase velocities
measured in the $a\rightarrow a$ and $b\rightarrow b$
transmission channels to determine $\chi_3$.
The transmitted amplitudes follow from the propagation
of the two eigenmodes:
\begin{align}
A_{a\leftarrow a}
&=e^{i(q_++q_-)L/2}
\left[
\cos\left(\frac{\delta qL}{2}\right)
+i\frac{\chi_3}
{\sqrt{\chi_3^2+\chi_1^2+(\omega\eta)^2}}
\sin\left(\frac{\delta qL}{2}\right)
\right],\\
A_{b\leftarrow b}
&=e^{i(q_++q_-)L/2}
\left[
\cos\left(\frac{\delta qL}{2}\right)
-i\frac{\chi_3}
{\sqrt{\chi_3^2+\chi_1^2+(\omega\eta)^2}}
\sin\left(\frac{\delta qL}{2}\right)
\right].
\end{align}
Taking the difference between their phases eliminates
the common contribution $(q_++q_-)L/2$, giving
\begin{equation}
\begin{aligned}
&\arg A_{a\leftarrow a}
-\arg A_{b\leftarrow b}
\\
&\quad =
2\tan^{-1}\left[
\frac{\chi_3}
{\sqrt{\chi_3^2+\chi_1^2+(\omega\eta)^2}}
\tan\left(\frac{\delta qL}{2}\right)
\right].
\end{aligned}
\end{equation}
The measured effective phase velocities satisfy
\begin{equation}
\begin{aligned}
\arg A_{a\leftarrow a}
-\arg A_{b\leftarrow b}
&=\omega L\left(\frac{1}{v_a}-\frac{1}{v_b}\right)\\
&=\frac{\omega L}{v_av_b}(v_b-v_a)\\
&\simeq\frac{\rho\omega L}{C_0}(v_b-v_a),
\end{aligned}
\end{equation}
where the last step uses $v_av_b\simeq C_0/\rho$,
valid to leading order in the spin-induced elastic
corrections.

Combining the two expressions gives
\begin{equation}
\frac{\rho\omega L}{2C_0}(v_b-v_a)
\simeq
\tan^{-1}\left[
\frac{\chi_3}
{\sqrt{\chi_3^2+\chi_1^2+(\omega\eta)^2}}
\tan\left(\frac{\delta qL}{2}\right)
\right].
\label{eq:dv_difference_SM}
\end{equation}
Taking the tangent and rearranging yields
\begin{equation}
\chi_3
\simeq
\sqrt{\chi_3^2+\chi_1^2+(\omega\eta)^2}\,
\frac{
\tan\left[
\frac{\rho\omega L}{2C_0}(v_b-v_a)
\right]
}{
\tan\left(\frac{\delta qL}{2}\right)
}.
\label{eq:delta_general_SM}
\end{equation}
Assuming $(\omega\eta)^2\ll\chi_3^2+\chi_1^2$
and $C_I\simeq C_0$, and using the independently determined ratio
$\chi_1/\chi_3$ from Eq.~\eqref{eq:chi_delta_SM},
we obtain
\begin{equation}
\begin{aligned}
\chi_3
\simeq{}&
\frac{2C_0}
{q_0L\sqrt{1+(\chi_1/\chi_3)^2}}
\tan^{-1}\left\{
\sqrt{1+(\chi_1/\chi_3)^2}
\tan\left[
\frac{\rho\omega L}{2C_0}(v_b-v_a)
\right]
\right\}.
\end{aligned}
\label{eq:delta_inferred_SM}
\end{equation}
Importantly, no expansion in $\delta qL$ has been made.

\section{Obtaining the spin excitations and gap}
In this section, we specify the spin model for $\alpha$-RuCl$_3$ and describe its magnon excitations.
We consider the generic $J-K-\Gamma-\Gamma'$ spin Hamiltonian introduced in Ref.~\cite{Rau2014PRL}, supplemented by the interlayer coupling described in Ref.~\cite{Cen2025PRB}:
\begin{align} 
       H_{\text{s}} = 
      &\sum_{\langle ij \rangle_{\parallel} \in \gamma} \left[ J \mathbf{s}_i \cdot \mathbf{s}_j + K s_i^\gamma s_j^\gamma + \Gamma (s_i^\alpha s_j^\beta + s_i^\beta s_j^\alpha) + \Gamma' (s_i^\gamma s_j^\alpha + s_i^\gamma s_j^\beta + s_i^\alpha s_j^\gamma + s_i^\beta s_j^\gamma) \right] \nonumber\\ 
        &\quad + \sum_{\langle ij \rangle_{\perp} \in \gamma} K_c s_i^\gamma s_j^\gamma
        - \mu_{\text{B}} \sum_{i} \mathbf{h} \cdot \mathbf{g} \cdot \mathbf{s}_{i}.
\end{align}
Here, $\mathbf{s}_i$ denotes the spin--orbit-entangled $j_{\rm eff}=1/2$ pseudospin operator at site $i\equiv\mathbf{R}_i$. The notation $\langle ij\rangle_{\parallel}$ and $\langle ij\rangle_{\perp}$ denotes in-plane and out-of-plane nearest-neighbor pairs, respectively. For an in-plane $\gamma$ bond ($\gamma\in\{x,y,z\}$), $\alpha$ and $\beta$ denote the remaining two orthogonal spin components. The in-plane interactions consist of the isotropic Heisenberg exchange $J$, the bond-dependent Kitaev interaction $K$, and the symmetric off-diagonal exchanges $\Gamma$ and $\Gamma'$, while $K_c$ denotes the interlayer Kitaev coupling which captures the two different zigzag magnetic patterns along the c-axis as shown in Ref. \cite{Cen2025PRB}. In the Zeeman term, $\mu_{\text{B}}$ is the Bohr magneton, $\mathbf{h}$ is the applied magnetic field, and $\mathbf{g}$ is the electronic $g$-tensor, which is diagonal in the absence of lattice strain.

We use the Julia package \texttt{Sunny}~\cite{Sunny_jl_2025}
to determine the classical magnetic ground state and the
corresponding spin-wave excitations.
For each field, the classical energy is minimized over the
spin orientations, starting from multiple random configurations
at zero field and subsequently minimizing as the field increases.
For a field applied in the $bc$ plane at an angle
$\theta=55^\circ$ from the $c$ axis, we consider the
$X$- and $Y$-zigzag domains.
The numerical calculation uses a $(1,2,1)$ magnetic
supercell containing 12 spins.
At sufficiently large fields, the classical spins become
field-polarized.

We then perform a Holstein--Primakoff expansion around
the classical spin configuration. At each sublattice $l$,
we introduce a local spin frame with its $z_l$ axis
aligned with the classical spin direction.
To leading order in the bosonic operators,
\begin{equation}
\begin{aligned}
s_i^{z_l}&=S-a_i^\dagger a_i,\\
s_i^{+_l}&\simeq\sqrt{2S}\,a_i,\\
s_i^{-_l}&\simeq\sqrt{2S}\,a_i^\dagger,
\end{aligned}
\end{equation}
where $S=1/2$.
Labelling the magnetic unit cells by $\mathbf R$
and the sublattices by $l$, we Fourier-transform
the bosonic operators as
\begin{equation}
a_{\mathbf R,l}^\dagger
=\frac{1}{\sqrt{N_{\rm cell}}}
\sum_{\mathbf k}
a_{\mathbf k,l}^\dagger
e^{i\mathbf k\cdot\mathbf R},
\end{equation}
where $N_{\rm cell}$ is the number of magnetic unit cells.
Retaining terms quadratic in the bosonic operators,
we diagonalize the resulting Hamiltonian through
a bosonic Bogoliubov transformation,
\begin{equation}
{\cal H}_{\rm spin}^0
\simeq E_{\rm SW}
+\sum_{\mathbf k,n}
E_{n,\mathbf k}\,
b_{n,\mathbf k}^\dagger b_{n,\mathbf k},
\end{equation}
where $E_{n,\mathbf k}$ are the magnon energies
and $n$ labels the magnon branches.

To calculate the responses in Eq.~(14) of the main text,
we expand the spin--phonon interactions
${\cal H}_{\rm int}^{(1)}$ and ${\cal H}_{\rm int}^{(2)}$,
derived in the following sections, in the same
Holstein--Primakoff bosons.
The terms linear in the bosonic operators generate
matrix elements between the ground state and
one-magnon states, while the quadratic terms generate
matrix elements involving two-magnon states.
The single-site interaction ${\cal H}_{\rm int}^{(1)}$
contributes to the one-magnon sector, whereas the
exchange interaction ${\cal H}_{\rm int}^{(2)}$
contributes to both sectors.

For an acoustic phonon with wave vector
$\mathbf q\parallel c$, the relevant excited states
are therefore the one-magnon states
$|n,\mathbf q\rangle$ and the two-magnon states
$|n,\mathbf k;m,\mathbf q-\mathbf k\rangle$,
with excitation energies
\begin{equation}
E_{n,\mathbf q},
\qquad
E_{n,\mathbf k}+E_{m,\mathbf q-\mathbf k},
\end{equation}
respectively.
Their matrix elements are obtained by expressing
$\partial_{\epsilon_\mu}{\cal H}_{\rm int}$
in terms of the Bogoliubov quasiparticles.
Both contributions are retained in the spectral sums
for $\chi_{\mu\nu}$ and $\Omega_{ba}$.
The two-magnon contribution is summed over the
magnetic Brillouin zone, with a factor of $1/2$
to avoid double counting the two-magnon pairs.
Numerically, the two-magnon response is evaluated
using a $21\times21\times5$ momentum mesh in the
magnetic Brillouin zone.

Finally, we evaluate the responses separately for
the $X$- and $Y$-zigzag domains.
The symmetric response tensor $\chi_{\mu\nu}$ is
rotated into the common crystallographic frame
before averaging the two domains with equal weights.
The antisymmetric response $\Omega_{ba}$ is invariant
under these in-plane proper rotations and is averaged
directly.
\section{Strain-induced off-diagonal {$g$}-tensor}

In this section we derive the strain-induced correction to the $g$-tensor in $\alpha$RuCl$_3$. We consider a magnetic ion in an octahedral ligand environment, where the dominant crystal field separates the $t_{2g}$ and $e_g$ manifolds. We restrict the analytic treatment to the low-energy $t_{2g}$ manifold and assume that both the Zeeman coupling and the strain-induced modification of the local crystal field are small compared with the spin--orbit coupling (SOC) scale. Within the $t_{2g}$ manifold, we use an effective orbital angular momentum $l_{\mathrm{eff}}=1$, whose operator is denoted by $\mathbf L$. SOC then produces a $j_{\mathrm{eff}}=1/2$ ground-state doublet and a $j_{\mathrm{eff}}=3/2$ excited-state quartet separated by an energy $3\lambda/2$.

Our goal is to obtain the coupling between strain and the low-energy $j_{\mathrm{eff}}=1/2$ doublet in the presence of an applied magnetic field. We focus first on the $\epsilon_{ca}$ strain component and write
\begin{align}
    H &= H_0 + H',\\
    H_0 &= H_{\mathrm{SOC}},\\
    H' &= H_{\mathrm{Zeeman}}+\delta V,
\end{align}
with
\begin{align}
    H_{\mathrm{Zeeman}}
    &= -\mu_B\mathbf h\cdot\left(-\mathbf L+2\mathbf S\right),\\
    \delta V
    &= \zeta_{ca}\epsilon_{ca}
    \left(L_cL_a+L_aL_c\right),
\end{align}
where $\mathbf S$ is the spin angular momentum, $\mathbf h=(h_a,h_b,h_c)$ is the applied magnetic field, and $\zeta_{ca}$ parametrizes the strain-induced quadrupolar crystal-field distortion.

We integrate out the excited $j_{\mathrm{eff}}=3/2$ manifold perturbatively. Let $P$ project onto the $j_{\mathrm{eff}}=1/2$ doublet and $Q=1-P$ onto the excited quartet. To second order in $H'$, the effective Hamiltonian within the low-energy manifold is
\begin{equation}
    H_{\mathrm{eff}}
    =
    P H_0 P
    +
    P H' P
    +
    P H' Q
    \frac{1}{E_0-QH_0Q}
    Q H' P,
\end{equation}
where $E_0$ is the energy of the $j_{\mathrm{eff}}=1/2$ doublet. Since the excited quartet lies higher in energy by $3\lambda/2$, the energy denominator is $-3\lambda/2$. The term $P H_{\mathrm{Zeeman}}P$ gives the usual Zeeman coupling within the $j_{\mathrm{eff}}=1/2$ manifold, while the terms linear in both magnetic field and strain arise from
\begin{equation}
\begin{aligned}
    H_{\mathrm{eff}}^{(h\epsilon)}
    =
    P H_{\mathrm{Zeeman}} Q
    \frac{1}{E_0-QH_0Q}
    Q\,\delta V\,P+
    P\,\delta V\,Q
    \frac{1}{E_0-QH_0Q}
    Q H_{\mathrm{Zeeman}}P.
\end{aligned}
\end{equation}
These cross-terms therefore describe the strain-induced modification of the $g$-tensor. Evaluating the matrix elements between the
$j_{\mathrm{eff}}=1/2$ doublet and the
$j_{\mathrm{eff}}=3/2$ quartet gives, in the
pseudospin basis quantized along the $c$ axis,
\begin{equation}
H_{\mathrm{eff}}^{(h\epsilon)}
=
-\frac{4\mu_B\zeta_{ca}\epsilon_{ca}}{3\lambda}
\begin{pmatrix}
h_a & h_c\\
h_c & -h_a
\end{pmatrix}.
\end{equation}
Identifying the Pauli matrices with twice the
pseudospin operators, the resulting spin-strain
interaction is
\begin{equation}
\mathcal{H}_{\mathrm{int}}^{(1)}
=
-\sum_i\mu_B\tilde g\,\epsilon_{ca}(\mathbf r_i)
\left(h_a s_i^c+h_c s_i^a\right)
+(a\leftrightarrow b),
\end{equation}
where $\tilde g=\frac{8\zeta_{ca}}{3\lambda}.$
We transform the real-space coupling to momentum space,
\begin{equation}
\begin{aligned}
    H_{s-p}^{(1)}
    ={}&
    -\mu_B\tilde g
    \sum_{\mathbf q,l}
    e^{i\mathbf q\cdot\mathbf r_l}
    \epsilon_{ca}(\mathbf q)
    \left[
        s_l^c(-\mathbf q)h_a
        +
        s_l^a(-\mathbf q)h_c
    \right]+(a\leftrightarrow b).
\end{aligned}
\end{equation}

To estimate the strain-induced crystal-field
coupling $\zeta_{ca}$, we consider a screened
electrostatic potential generated by the six
Cl ligands surrounding a Ru ion,
\begin{equation}
V_{\rm Cl}(\mathbf r)
=
\sum_n
\frac{Z_{\rm eff}
e^{-\kappa|\mathbf r-\mathbf R_n|}}
{|\mathbf r-\mathbf R_n|},
\end{equation}
where $Z_{\rm eff}$ and
$\kappa$ characterize the effective ligand
charge and screening, respectively.
We consider an idealized RuCl$_6$ geometry
with a Ru--Ru separation of $6.6$ Bohr
and a $5\%$ trigonal compression along
the $c$-axis.

The crystal-field matrix elements are
evaluated using the Clementi--Roetti
Ru $4d$ atomic orbitals
\cite{Clementi1974JCP}.
To introduce $\epsilon_{ca}$, we displace
the Cl ions according to
$u_a=\epsilon_{ca}z$, where $z$ denotes
their position along the $c$-axis relative
to the central Ru ion, which remains fixed.
We then extract $\zeta_{ca}$ from the
linear strain dependence of the crystal-field
Hamiltonian in the one-hole representation.

For the parameters
$Z_{\rm eff}=0.7$ and $\kappa r_0=2$,
where $r_0$ is the equilibrium Ru--Cl
distance, we obtain
$\zeta_{ca}\simeq0.411$~eV and
$\zeta_{bc}\simeq0.411$~eV.
Taking $\lambda=0.176$~eV gives
$\tilde g\simeq6.2$.
The screening parameters are chosen
phenomenologically, and the estimate
illustrates the electrostatic contribution
to the strain-induced $g$-tensor.
\section{Strain-modified exchange interactions}

To estimate the strain-induced changes in the spin-exchange interactions, we examine their microscopic origin in the underlying hopping processes. We focus first on the $z$-bond, which lies parallel to the crystallographic $b$-axis. Starting from a multi-orbital Kanamori Hamiltonian in the strong-coupling regime, $U,J_H\gg\lambda\gg t$, where $U$ is the on-site Coulomb repulsion and $J_H$ is the Hund's coupling, the effective hopping matrix on the $z$-bond in the basis $\{\ket{yz},\ket{zx},\ket{xy}\}$ is
\begin{equation}
T=
\begin{pmatrix}
        t_1^{yz} & t_2 & t_4^{yz} \\
        t_2 & t_1^{zx} & t_4^{zx} \\
        t_4^{yz} & t_4^{zx} & t_3
\end{pmatrix}.
\end{equation}
In the unstrained lattice, symmetry imposes
$t_4^{zx}=t_4^{yz}\equiv t_4$ and
$t_1^{zx}=t_1^{yz}\equiv t_1$.
Following the approach in Ref. \cite{Rau2014PRL} we find the effective spin-exchange interactions to leading order in $J_H/U$:
\begin{align}
K &= \frac{8}{9U}\left(\frac{J_H}{U}\right)
\left[
(-t_1^{yz}+t_3)(-2t_1^{zx}+t_1^{yz}+t_3)
-3(t_2^2-{t_4^{zx}}^2)
\right],\\
K' &= \frac{8}{9U}\left(\frac{J_H}{U}\right)
\left[
(t_1^{zx}-t_1^{yz})(t_1^{zx}+t_1^{yz}-2t_3)
+3(t_4^{zx}-t_4^{yz})(t_4^{zx}+t_4^{yz})
\right],\\
\Gamma &= \frac{8}{9U}\left(\frac{J_H}{U}\right)
\left[
3t_4^{zx}t_4^{yz}
+t_2(t_1^{zx}+t_1^{yz}-2t_3)
\right],\\
\Gamma' &= -\frac{8}{9U}\left(\frac{J_H}{U}\right)
\left[
t_4^{yz}(2t_1^{zx}-t_1^{yz}-t_3)
-3t_2t_4^{zx}
\right],\\
\Gamma'' &= -\frac{8}{9U}\left(\frac{J_H}{U}\right)
\left[
t_4^{zx}(2t_1^{yz}-t_1^{zx}-t_3)
-3t_2t_4^{yz}
\right].
\end{align}
The generalized spin interaction on the $z$-bond is therefore described by the exchange matrix
\begin{equation}
    \begin{pmatrix}
        J & \Gamma & \Gamma'\\
        \Gamma & J+K' & \Gamma''\\
        \Gamma' & \Gamma'' & J+K
    \end{pmatrix}.
\end{equation}

Transforming to the global crystallographic frame,
the $ca$ and $bc$ spin interactions along an
in-plane bond $\gamma$ can be parametrized as
\begin{equation}
H_{ca/bc}^{\gamma}
=
-\sqrt{2}J_{\pm c}
\left[
{\rm c}_\gamma
(s_i^a s_j^c+s_i^c s_j^a)
+
{\rm s}_\gamma
(s_i^b s_j^c+s_i^c s_j^b)
\right],
\end{equation}
where ${\rm c}_\gamma=\cos\phi_\gamma$ and
${\rm s}_\gamma=\sin\phi_\gamma$.
For the $z$-bond, transforming the generalized
exchange matrix gives
\begin{align}
J_{\pm c}\cos\phi_z
&=
-\frac{1}{6}
\left(
2\Gamma-\Gamma'-\Gamma''+K'-2K
\right),\\
J_{\pm c}\sin\phi_z
&=
-\frac{1}{2\sqrt{3}}
\left(
\Gamma''-\Gamma'+K'
\right).
\end{align}
These relations determine both the magnitude
$J_{\pm c}$ and the orientation $\phi_z$
of the interaction in terms of the microscopic
exchange parameters.

In the unstrained limit, $K'=0$ and
$\Gamma''=\Gamma'$, giving
\begin{equation}
J_{\pm c}=\frac{K-\Gamma+\Gamma'}{3},
\qquad
\phi_z=0.
\end{equation}
The corresponding bond angles are
$\phi_\gamma=0,2\pi/3,4\pi/3$ for the
$z$-, $x$-, and $y$-bonds, respectively,
recovering the interaction introduced
in the main text.

We next determine how out-of plane shear strain modifies these interactions. For transverse acoustic phonons propagating along
the $c$-axis, the two Ru sites on an in-plane bond undergo the same displacement, while the out-of-plane Cl ligands move relative to them. The direct Ru--Ru hopping therefore does not change linearly with strain, and the dominant response arises from Ru--Cl--Ru pathways. For the $z$-bond, $\epsilon_{ca}$ modifies the magnitude $J_{\pm c}$, whereas $\epsilon_{bc}$ induces a change in the angle $\phi_z$. By symmetry, and as is found numerically below, the corresponding linear changes of $J_{\pm c}$ under $\epsilon_{bc}$ and of $\phi_z$ under $\epsilon_{ca}$ vanish. The corresponding magnetoelastic coupling associated with the change in $J_{\pm c}$ is
\begin{equation}
g_1
\equiv
-\sqrt{2}
\left.
\frac{\partial J_{\pm c}}
{\partial\epsilon_{ca}}
\right|_{z-bond}=
\frac{1}{3\sqrt{2}}
\left.
\frac{\partial}
{\partial\epsilon_{ca}}
\left(
2\Gamma-\Gamma'-\Gamma''+K'-2K
\right)
\right|_{z-bond}.
\end{equation}
By contrast, $\epsilon_{bc}$ breaks the local
symmetry between the off-diagonal exchange channels, allowing $\Gamma'$ and $\Gamma''$ to become inequivalent. This changes the angle $\phi_z$, generating an $s_i^b s_j^c+s_i^c s_j^b$ interaction. The associated magnetoelastic coupling is
\begin{equation}
g_2
\equiv
-\sqrt{2}J_{\pm c}
\left.
\frac{\partial\phi_\gamma}
{\partial\epsilon_{bc}}
\right|_{z-bond}=
\frac{1}{\sqrt{6}}
\left.
\frac{\partial}
{\partial\epsilon_{bc}}
\left(
\Gamma''-\Gamma'+K'
\right)
\right|_{z-bond}.
\end{equation}

The resulting strain-induced interaction on the $z$-bond is
\begin{equation}
   {\cal H}_{\rm int}^{(2),z}
   =
   g_1\epsilon_{ca}(\mathbf r_{ij})
   (s_i^a s_j^c+s_i^c s_j^a)
   +
   g_2\epsilon_{bc}(\mathbf r_{ij})
   (s_i^b s_j^c+s_i^c s_j^b),
\end{equation}
where $\mathbf r_{ij}$ denotes the center of the bond
connecting sites $i$ and $j$. For transverse acoustic
phonons propagating along the $c$-axis, the strain
depends only on the $c$ coordinate. Since the two
sites of an in-plane bond lie in the same $ab$ plane,
$\epsilon_{\mu}(\mathbf r_{ij})
=\epsilon_{\mu}(\mathbf r_i)
=\epsilon_{\mu}(\mathbf r_j)$,
consistent with the convention used in the main text.

The corresponding interactions on the other in-plane
bonds follow by rotating both the in-plane spin
components and the shear strains by the bond angle
$\phi_\gamma$. This produces the coefficients
$g_1{\rm c}_\gamma^2+g_2{\rm s}_\gamma^2$
and $g_1{\rm s}_\gamma^2+g_2{\rm c}_\gamma^2$
for the diagonal couplings, and
$(g_1-g_2){\rm s}_\gamma{\rm c}_\gamma$
for the mixed terms. Summing over all in-plane bonds gives
\begin{equation}
    {\cal H}_{\rm int}^{(2)}=
    \sum_{\langle ij\rangle_\parallel\in\gamma}
    \epsilon_{ca}(\mathbf r_{ij})
    \Big\{
    (g_1{\rm c}_\gamma^2+g_2{\rm s}_\gamma^2)
    (s_i^a s_j^c+s_i^c s_j^a)
    +{\rm s}_\gamma{\rm c}_\gamma(g_1-g_2)
    (s_i^b s_j^c+s_i^c s_j^b)
    \Big\}
    +(a\leftrightarrow b,\,
    {\rm c}_\gamma\leftrightarrow{\rm s}_\gamma).
\end{equation}

This interaction can then be expressed in the
spin-wave basis introduced above to evaluate the
matrix elements involving one- and two-magnon
excitations that enter the static and geometric
responses.

To estimate the magnetoelastic couplings, we calculate the strain-dependent hopping amplitudes from orbital overlaps using Clementi--Roetti Ru $4d$ and Cl $3p$ orbitals \cite{Clementi1974JCP} within an extended H\"uckel approach. We use the Wolfsberg--Helmholtz prescription with a parameter of $1.75$ \cite{Wolfsberg1952JCP,Hoffmann1963JCP} and representative orbital energies of $-0.454$ and $-0.522$ Hartree for Ru $4d$ and Cl $3p$, respectively \cite{Cerda200PRB,Dedieu1979JACS}. The direct Ru--Ru hopping is kept fixed under strain, while the ligand-mediated Ru--Cl--Ru hopping varies with strain. Substituting the resulting hopping amplitudes into the exchange expressions above, we obtain $\frac{g_1}{J_{\pm c}}\simeq 9.8$. The estimates depend on the orbital energies and ligand geometry, which affect the relative contributions of direct and ligand-mediated hopping. While $g_1$ is of order $10J_{\pm c}$, we find that $g_2$ is generally smaller, of order $J_{\pm c}$. Its sign is particularly sensitive to the microscopic parameters because it involves competing strain derivatives of $\Gamma'$, $\Gamma''$, and $K'$. Indeed, varying the orbital energies and ligand geometry can reverse its sign. We therefore regard the value of $g_2$ as an order-of-magnitude estimate and treat its sign phenomenologically in the spin-wave calculation.

\bibliographystyle{apsrev4-2}
\bibliography{cites}